\documentclass[aps,pre,twocolumn,showpacs,superscriptaddress]{revtex4-1}
\usepackage{bm}
\usepackage{mathrsfs}
\usepackage{amsmath}
\usepackage{amssymb}
\usepackage{graphicx}
\usepackage{amsfonts}
\usepackage{amsthm}
\usepackage{color}
\usepackage{dcolumn}
\usepackage{txfonts}
\usepackage{epsfig}

\begin{document}

\title{Quantum Work Relations and Response Theory in $\mathcal{PT}$-Symmetric Quantum Systems}
\author{Bo-Bo Wei}
\affiliation{School of Physics and Energy, Shenzhen University, 518060 Shenzhen, China}

\begin{abstract}
In this work, we show that a universal quantum work relation for a quantum system driven arbitrarily far from equilibrium extend to $\mathcal{PT}$-symmetric quantum system with unbroken $\mathcal{PT}$ symmetry, which is a consequence of microscopic reversibility. The quantum Jarzynski equality, linear response theory and Onsager reciprocal relations for the $\mathcal{PT}$-symmetric quantum system are recovered as special cases of the universal quantum work relation in $\mathcal{PT}$-symmetric quantum system. In the regime of broken $\mathcal{PT}$ symmetry, the universal quantum work relation does not hold as the norm is not preserved during the dynamics.
\end{abstract}
\maketitle

\section{Introduction}
In 1997, Jarzynski \cite{Jarzynski1997} made a remarkable development by discovering that a classical
system prepared in the canonical equilibrium state the work done on the classical system under variation of an externally control parameter
is connected to the Helmholtz free energy differences between the initial and the final equilibrium states for the
control parameters. The Jarzynski equality establishes connections between the equilibrium free energy difference and the work in a non-equilibrium process \cite{PNAS2001,Science2005,Nature2005,EPL2005,PT2005,PRL2006,PRL2007} and thus provide a novel method to study thermodynamics in nanscale sytems by non-equilibrium measurement \cite{PNAS2001,Science2005,Nature2005,EPL2005,PT2005,PRL2006,PRL2007}. Jarzynski equality was generalized to finite quantum mechanical systems \cite{arXiv2000a,arXiv2000b,PRL2003,Talker2007} and has been verified experimentally in an ion trap system\cite{Kim2015}. The finding of the Jarzynski equality has generated extensive investigation of fluctuation relations in non-equlibrium thermodynamics \cite{RMP2009,Jar2011,RMP2011,Wei2017a}.

Recently the Jarzynski equality and some fluctuation relations were generalized to the $\mathcal{PT}$-symmetric quantum systems \cite{PT2015,PT2016,PT2017,PT2017b}. The motivation of the present work is to investigate whether the universal quantum work relation \cite{workrelation2008} which connects non-equilibrium measurement in the forward process and that in its the time reversed process survives in the $\mathcal{PT}$ symmetric quantum systems? We found that this is indeed the case for $\mathcal{PT}$-symmetric quantum systems in the phase of unbroken $\mathcal{PT}$ symmetry. On the other hand, for $\mathcal{PT}$-symmetric quantum systems which is in the phase of broken $\mathcal{PT}$ symmetry, the universal quantum work relation does not hold.

The rest of this paper is organized as follows: In Sec.~II, we briefly review the mathematical formalism of time-dependent $\mathcal{PT}$-symmetric quantum mechanics. In Sec.~III, we define the forward and the reversed processes of non-equilibrium thermodynamics for $\mathcal{PT}$-symmetric quantum systems and then derive the universal work relations for $\mathcal{PT}$-symmetric quantum systems. In Sec. IV, we summarize our findings.

\section{Formalism of $\mathcal{PT}$-symmetric Quantum Systems}
In this section, we give a brief review on the central features of $\mathcal{PT}$-symmetric quantum mechanics. For non-Hermitian Hamiltonian \cite{NonHermitian2011}, i.e. $\mathcal{H}\neq\mathcal{H}^{\dagger}$, the left eigenvectors and right eigenvectors are usually different and the Schr\"{o}dinger equation for them are respectively,
\begin{eqnarray}
\mathcal{H}|\psi_n\rangle&=&E_n|\psi_n\rangle,\\
\langle \phi_n|\mathcal{H}&=&E_n\langle \phi_n|,
\end{eqnarray}
where $n$ is the quantum number that characterizes different eigenstate and $E_n$ is the eigenenergy of the corresponding state $|\psi_n\rangle$ and we assume the eigenstates of $\mathcal{H}$ are discrete. Note that the left eigenvector and the corresponding right eigenvector share the same eigenenergy \cite{NonHermitian2011}. The eigenstates of non-Hermitian Hamiltonian is not normalized as that of Hermitian quantum mechanics while the left eigenvector and the corresponding right eigenvector are normalized, namely $\langle\psi_n|\phi_m\rangle=\delta_{mn}$ and $\sum_n|\psi_n\rangle\langle\phi_n|=1$. If a non-Hermitian Hamiltonian $\mathcal{H}$ has $\mathcal{PT}$ symmetry and also locates in the phase of unbroken $\mathcal{PT}$ symmetry, then there exists $\mathcal{G}$ such that \cite{PesudoHermitian1,PesudoHermitian2,PesudoHermitian3}
\begin{eqnarray}
\mathcal{H}^{\dagger}=\mathcal{G}\mathcal{H}\mathcal{G}^{-1},
\end{eqnarray}
where $\mathcal{G}$ and $\mathcal{G}^{-1}$ are respectively \cite{PesudoHermitian1,PesudoHermitian2,PesudoHermitian3}
\begin{eqnarray}
\mathcal{G}&=&\sum_n|\phi_n\rangle\langle\phi_n|,\\
\mathcal{G}^{-1}&=&\sum_n|\psi_n\rangle\langle\psi_n|.
\end{eqnarray}
So for an arbitrary state $|\Phi\rangle$ the normalization condition should be written as \cite{PesudoHermitian1,PesudoHermitian2,PesudoHermitian3}
\begin{eqnarray}
\langle\Phi|\mathcal{G}|\Phi\rangle=1.
\end{eqnarray}
With $\mathcal{G}$ operator, the completeness relation becomes \cite{PesudoHermitian1,PesudoHermitian2,PesudoHermitian3}
\begin{eqnarray}
\sum_m|\phi_m\rangle\langle\phi_m|\mathcal{G}=1.
\end{eqnarray}

Now we consider the dynamics of a quantum state governed by a time-dependent Hamiltonian $\mathcal{H}(t)$ with $\mathcal{PT}$ symmetry. For $\mathcal{PT}$-symmetric quantum systems, the time development of a quantum state is governed by the Schr\"{o}dinger equation with modification in order to preserve unitarity \cite{PTevolution}. The modified Schr\"{o}dinger equation is \cite{PTevolution}
\begin{eqnarray}\label{SE}
i\partial_t|\psi\rangle=[\mathcal{H}(t)+\mathcal{A}(t)]|\psi\rangle.
\end{eqnarray}
Here the time dependent gauge field, $\mathcal{A}(t)=-i\mathcal{G}_t^{-1}\partial_t\mathcal{G}_t$, has been introduced to guarantee unitarity of the quantum dynamics in $\mathcal{PT}$-symmetric systems in the phase of unbroken $\mathcal{PT}$ symmetry. Then the time evolution operator is given by
\begin{eqnarray}\label{evolution}
\mathcal{U}_{0,t}=\mathcal{T}e^{-i\int_0^tdt'[\mathcal{H}(t')+\mathcal{A}(t')]},
\end{eqnarray}
where $\mathcal{T}$ is the time ordering operator which comes from the time-dependent Hamiltonian in the Schr\"{o}dinger equation, Equation \eqref{SE}. If the quantum system is in the unbroken $\mathcal{PT}$-symmetric phase, all the eigenvalues are real and the quantum dynamics generated by $\mathcal{U}_{0,t}$ is unitary. However, the time evolution operator $\mathcal{U}_{0,t}$ is not unitary but satisfies the relation\cite{PT2017b}
\begin{eqnarray}\label{unitarity}
\mathcal{U}_{0,t}^{\dagger}\mathcal{G}_{t}\mathcal{U}_{0,t}=\mathcal{G}_0.
\end{eqnarray}
This is the corresponding unitarity condition in $\mathcal{PT}$-symmetric quantum mechanics.

To study thermodynamics for $\mathcal{PT}$-symmetric systems,  some mathematical operations have to be changed to match the theory of statistical mechanics. Firstly, the inner product in the $\mathcal{PT}$-symmetric quantum systems has to be modified to be
\begin{eqnarray}
\langle\phi_1|\phi_2\rangle\rightarrow\langle\phi_1|\mathcal{G}|\phi_2\rangle.
\end{eqnarray}
Secondly, the trace operation has to be changed \cite{NonHermitian2011,PTevolution}
\begin{eqnarray}
\text{Tr}_{\mathcal{G}}[\mathcal{O}]=\sum_n\langle\phi_n|\mathcal{GO}|\phi_n\rangle,
\end{eqnarray}
where $\mathcal{O}$ is an arbitrary operator and $\{|\phi_n\rangle\}$ form a complete basis in the Hilbert space of the $\mathcal{PT}$-symmetric system. The modified trace operation satisfies the cyclic property \cite{PT2017}, \begin{eqnarray}\label{tr1}
\text{Tr}_{\mathcal{G}}[\mathcal{PQ}]=\text{Tr}_{\mathcal{G}}[\mathcal{QP}].
\end{eqnarray}
Here $\mathcal{P}$ and $\mathcal{Q}$ are two arbitrary operators.

\section{Non-equilibrium Quantum thermodynamics for $\mathcal{PT}$-symmetric Systems}
For a non-Hermitian Hamiltonian $\mathcal{H}$ with $\mathcal{PT}$ symmetry, the quantum state generally presents two different phases. One is the phase of unbroken $\mathcal{PT}$ symmetry in which all the eigenvalues are real. The other is the phase of broken $\mathcal{PT}$ symmetry in which the eigenvalues of the Hamiltonian have real and imaginary parts. The universal critical behaviors in non-Hermitian phase transitions have been recently studied \cite{Wei2017b}. Thermodynamics in the complex plane of control parameters have been extensively investigated through the quantum decoherence of a probe spin
\cite{Wei2012,Wei2014,Wei2015,Peng2015,LYExp2015,Wei2017a1,Wei2017b1} and large derivation statistics \cite{DynamicLY2013,DynamicLY2017}.

We consider a finite quantum system with a $\mathcal{PT}$ symmetric Hamiltonian $\mathcal{H}$ and we are interested in a non-equilibrium process of the $\mathcal{PT}$-symmetric system defined by a time-dependent parameter $\lambda(t)$ which is controlled by an external agent. In the following, we consider a driving protocol for which the Hamiltonian $\mathcal{H}(t)\equiv\mathcal{H}(\lambda(t))$ is in the phase of unbroken $\mathcal{PT}$ symmetry for all times so that the eigenvalues of $\mathcal{H}(t)$ are all real and the quantum dynamics are unitary.

\subsection{Forward Process for $\mathcal{PT}$-symmetric Systems}
Let us first define the \emph{forward non-equilibrium process in a $\mathcal{PT}$-symmetric quantum system} under time-dependent driving. At initial time $t=0$, we initialize the $\mathcal{PT}$-symmetric quantum system in the canonical equilibrium state with inverse temperature $\beta=1/T$ ($k_B=1$) at a fixed value of control parameter $\lambda_i$, which is given by
\begin{eqnarray}\label{forward0}
\rho_F(0)=e^{-\beta \mathcal{H}(\lambda_i)}/Z(\beta,\lambda_i),
\end{eqnarray}
where $Z(\beta,\lambda_i)=\text{Tr}_{\mathcal{G}}[e^{-\beta \mathcal{H}(\lambda_i)}]$ is the canonical partition function of the initial equilibrium state. Then we isolate the $\mathcal{PT}$-symmetric system and drive it by the Hamiltonian $\mathcal{H}(\lambda(t))$ for a time interval $\tau$, where the drving protocol $\lambda(t), t\in[0,\tau]$ takes the parameter from $\lambda_i$ at $t=0$ to $\lambda_f$ at time $\tau$. Hence the quantum state at $t$ due to the driving protocol is given by \cite{PT2017b}
\begin{eqnarray}\label{ef}
\rho_F(t)&=&\mathcal{U}_{0,t}\rho_F(0)\mathcal{U}_{0,t}^{-1},
\end{eqnarray}
where $\mathcal{U}_{0,t}\equiv \mathcal{T}e^{-i\int_0^tdt'[\mathcal{H}(\lambda(t'))+\mathcal{A}(t')]}$ with $\mathcal{T}$ being the time ordering operator.  It is obvious that the non-equilibrium state $\rho(\tau)$ is different from the equilibrium state at the final control parameter, i.e. $\rho_f=e^{-\beta \mathcal{H}(\lambda_f)}/Z(\beta,\lambda_f)$.

In the above driving process, we do work on the system. The work done on a quantum system is defined by two projective measurements \cite{Talker2007,RMP2011}. For simplicity of notation, we assume $\mathcal{H}(\lambda)$ satisfies eigenvalues equation, $\mathcal{H}(\lambda)|n(\lambda)\rangle=E_n(\lambda)|n(\lambda)\rangle$ for any $\lambda$, where the quantum number $n$ labels different eigenvectors. At $t=0$, the first projective measurement of $\mathcal{H}(\lambda_i)$ is performed and the result is $E_n(\lambda_i)$ with the corresponding probability,
\begin{eqnarray}
p_n(0)&=&\text{Tr}_{\mathcal{G}_0}\left[\rho_F(0)|n(\lambda_i)\rangle\langle n(\lambda_i)|\mathcal{G}_0\right],\nonumber\\
&=&e^{-\beta E_n(\lambda_i)}/Z(\beta,\lambda_i).
\end{eqnarray}
At the same time, the initial equilibrium state $\rho_F(0)$ is projected into the corresponding eigen state $|n(\lambda_i)\rangle$ of $\mathcal{H}(\lambda_i)$ with eigenenergy $E_n(\lambda_i)$. In the time interval, $0<t<\tau$, the quantum system is driven by the evolution operator $\mathcal{U}_{0,\tau}$ and then the quantum state at time $\tau$ is $\mathcal{U}_{0,\tau}|n(\lambda_i)\rangle$. Finally at time $t=\tau$, the second projective measurement of $\mathcal{H}(\lambda_f)$ is performed and the outcome is $E_m(\lambda_f)$ with conditional probability,
\begin{eqnarray}
p_{n\rightarrow m}=|\langle m(\lambda_f)|\mathcal{G}_{\tau}\mathcal{U}_{0,\tau}|n(\lambda_i)\rangle|^2.
\end{eqnarray}
In the two projective measurements of energy, the work done on the system is $W=E_m(\lambda_F)-E_n(\lambda_i)$. While the probability of obtaining $W$ is equal to the probability of obtaining $E_n(\lambda_i)$ for the first measurement and followed by getting $E_m(\lambda_f)$ in the second measurement, which is $p_n(0)p_{n\rightarrow m}$. Thus the quantum work distribution is \cite{Talker2007,RMP2011}
\begin{eqnarray}\label{QWD}
P(W)=\sum_{m,n}p_n(0)p_{n\rightarrow m}\delta\left(W-E_m(\lambda_f)+E_n(\lambda_i)\right).
\end{eqnarray}
The characteristic function of quantum work distribution, which is the Fourier transformation of the quantum work distribution, is given by \cite{Talker2007,RMP2011}
\begin{eqnarray}\label{charac}
G(u)&=&\int_{-\infty}^{\infty}dWP(W)e^{iuW},\\
&=&Z(\beta,\lambda_i)^{-1}\text{Tr}_{\mathcal{G}_0}[\mathcal{U}_{0,\tau}e^{-(\beta+iu)\mathcal{H}(0)}\mathcal{U}_{0,\tau}^{-1}e^{iu\mathcal{H}(\tau)}].
\end{eqnarray}

\subsection{The Reversed Process for $\mathcal{PT}$-symmetric Systems}
Now we define the reversed process for $\mathcal{PT}$-symmetric quantum system. In the reversed process, we first initialize the quantum system in the time reversal of the thermodynamic equilibrium state at inverse temperature $\beta=1/T$ at control parameter $\lambda_f$, which is
\begin{eqnarray}\label{reversed1}
\rho_R(0)=\frac{\Theta e^{-\beta \mathcal{H}(\lambda_f)}\Theta^{-1}}{Z(\beta,\lambda_f)},
\end{eqnarray}
where $Z(\beta,\lambda_f)=\text{Tr}_{\mathcal{G}}[e^{-\beta\mathcal{H}(\lambda_f)}]$ is the canonical
partition function. Then we drive the system in the reversed process for a time interval $\tau$ by the Hamiltonian \cite{Stra1994},
\begin{eqnarray}
\mathcal{H}_R(t)=\Theta \mathcal{H}(\tau-t)\Theta^{-1}.
\end{eqnarray}
The time evolution operator in the reversed process is
\begin{eqnarray}
\mathcal{V}_{0,\tau}=\mathcal{T}e^{-i\int_0^{\tau}dt[\Theta(\mathcal{H}(\tau-t)+\mathcal{A}(\tau-t))\Theta^{-1}]}.
\end{eqnarray}
So the time development of the density operator in the reversed process is
\begin{eqnarray}\label{er}
\rho_R(t)&=&\mathcal{V}_{0,t}\rho_R(0)\mathcal{V}_{0,t}^{-1}.
\end{eqnarray}
Having defined the forward process and the time reversed process for $\mathcal{PT}$-symmetric quantum system, we are now ready to establish the quantum work relations for $\mathcal{PT}$-symmetric quantum system.

\subsection{The Quantum Work Relations in $\mathcal{PT}$-symmetric Systems}
First, we establish a relation, which connects the time evolution operator in the forward process $\mathcal{U}_{0,t}$ and the time evolution operator in the time reversed process $\mathcal{V}_{0,t}$ by the following\\
\emph{\textbf{Lemma:} The time evolution operators in the forward process and in its time reversed process for $\mathcal{PT}$-symmetric quantum system are related by}
\begin{eqnarray}\label{frrelation}
\mathcal{V}_{0,t}=\Theta\mathcal{U}_{0,\tau-t}\mathcal{U}_{0,\tau}^{-1}\Theta^{-1},
\end{eqnarray}
\emph{where $t$ is an arbitrary time, $0\leq t\leq \tau$.}\\
\textbf{Proof:}
First, we know that the quantum states $\rho_F(\tau)$ and $\rho_F(\tau-t)$ in the forward process are related by
\begin{eqnarray}\label{fo}
\rho_F(\tau)=\mathcal{U}_{\tau-t,\tau}\rho_F(\tau-t)\mathcal{U}_{\tau-t,\tau}^{-1}.
\end{eqnarray}
If we do a time reversal operation on the final state of the forward process $\rho_F(\tau)$, then we get $\Theta\rho_F(\tau)\Theta^{-1}$. Then we drive this time reversed state by the evolution operator in the reversed process, $\mathcal{V}_{0,t}$, for a time duration $t$. According to time reversal symmetry, the final evolved state should be the time reversed state of $\rho_F(\tau-t)$, i.e. $\Theta\rho_F(\tau-t)\Theta^{-1}$. This gives us
\begin{eqnarray}
\Theta\rho_F(\tau-t)\Theta^{-1}=\mathcal{V}_{0,t}\Theta\rho_F(\tau)\Theta^{-1}\mathcal{V}_{0,t}^{-1}.
\end{eqnarray}
Making use of Equation \eqref{fo}, we get
\begin{eqnarray}
\Theta\rho_F(\tau-t)\Theta^{-1}=\mathcal{V}_{0,t}\Theta\mathcal{U}_{\tau-t,\tau}\rho_F(\tau-t)U_{\tau-t,\tau}^{-1}\Theta^{-1}\mathcal{V}_{0,t}^{-1}.
\end{eqnarray}
Thus we obtain
\begin{eqnarray}
\Theta=\mathcal{V}_{0,t}\Theta\mathcal{U}_{\tau-t,\tau}.
\end{eqnarray}
This leads to,
\begin{eqnarray}\label{a1}
\mathcal{V}_{0,t}=\Theta\mathcal{U}_{\tau-t,\tau}^{-1}\Theta^{-1}.
\end{eqnarray}
Making use of the equality, $\mathcal{U}_{\tau-t,\tau}\mathcal{U}_{0,\tau-t}=\mathcal{U}_{0,\tau}$, we get
\begin{eqnarray}\label{a2}
\mathcal{U}_{\tau-t,\tau}^{-1}=\mathcal{U}_{0,\tau-t}\mathcal{U}_{0,\tau}^{-1}.
\end{eqnarray}
Combing Equations \eqref{a1} and Equation \eqref{a2}, we obtain
\begin{eqnarray}
\mathcal{V}_{0,t}=\Theta\mathcal{U}_{0,\tau-t}\mathcal{U}_{0,\tau}^{-1}\Theta^{-1}.
\end{eqnarray}
Thus the Lemma is proved.

With this Lemma, we are now to prove the following\\
\emph{\textbf{Theorem:} Let us consider an arbitrary time-independent observable $A$ with a definite parity under time reversal $\Theta A\Theta^{-1}=\epsilon_AA$ with $\epsilon_A=\pm1$. It satisfies the following functional relation:}
\begin{eqnarray}\label{ca}
&&\left\langle \exp\left(\int_0^{\tau}dt\lambda(t)A_F(t)\right) e^{-\beta \mathcal{H}_F(\tau)}e^{\beta \mathcal{H}(0)}\right\rangle_F,\nonumber \\
&=&e^{-\beta\Delta F}\left\langle \exp\left(\int_0^{\tau}\epsilon_A A_R(t)\lambda(\tau-t)dt\right)\right\rangle_R,
\end{eqnarray}
\emph{where $\lambda(t)$ is an arbitrary function and $F$ and $R$ denote the forward and reversed process for the $\mathcal{PT}$ symmetric quantum systems defined in Sec.~IIIA and Sec.~IIIB respectively. $\Delta F=F(\beta,\lambda_f)-F(\beta,\lambda_i)$ is the free energy difference between the equilibrium states at the initial and final control parameters. }\\
\textbf{Proof:}
Let us first consider the quantity $A_F(t)$, which is Heisenberg representation of $A$ and defined by
\begin{eqnarray}
A_F(t)&=&\mathcal{U}_{0,t}^{-1}A\mathcal{U}_{0,t}.
\end{eqnarray}
Making use of the Lemma, Equation \eqref{frrelation}, we get
\begin{eqnarray}
A_F(t)&=&\mathcal{U}_{0,\tau}^{-1}\Theta^{-1}\mathcal{V}_{0,\tau-t}^{-1}\Theta A \Theta^{-1}\mathcal{V}_{0,\tau-t}\Theta\mathcal{U}_{0,\tau} ,\\
&=&\epsilon_A\mathcal{U}_{0,\tau}^{-1}\Theta^{-1}\mathcal{V}_{0,\tau-t}^{-1}A\mathcal{V}_{0,\tau-t}\Theta\mathcal{U}_{0,\tau} ,\\
&=&\epsilon_A \mathcal{U}_{0,\tau}^{-1}\Theta^{-1}A_R(\tau-t)\Theta \mathcal{U}_{0,\tau}.
\end{eqnarray}
Integrating over time with an arbitrary function $\lambda(t)$ and taking the exponential on both sides of the above equation, we then obtain,
\begin{eqnarray}
&&\exp\left(\int_0^{\tau}dt\lambda(t)A_F(t)\right),\nonumber \\&=&\mathcal{U}_{0,\tau}^{-1}\Theta^{-1}\exp\left(\int_0^{\tau}\epsilon_A A_R(\tau-t)\lambda(t)dt\right)\Theta \mathcal{U}_{0,\tau},\label{d1}\\
&=&\mathcal{U}_{0,\tau}^{-1}\Theta^{-1}\exp\left(\int_0^{\tau}\epsilon_A A_R(t)\lambda(\tau-t)dt\right)\Theta \mathcal{U}_{0,\tau}. \label{d2}
\end{eqnarray}
From Equation \eqref{d1} to Equation \eqref{d2}, we just change variable for integration. Now we are ready to prove the final equality for work relations and we start from the left side of the equation \eqref{ca},
\begin{eqnarray}
&&\left\langle \exp\left(\int_0^{\tau}dt\lambda(t)A_F(t)\right) e^{-\beta \mathcal{H}_F(\tau)}e^{\beta \mathcal{H}(0)}\right\rangle_F,\nonumber\\
&=&\text{Tr}_{\mathcal{G}}\left[\rho_F(0)e^{\int_0^{\tau}dt\lambda(t)A_F(t)} e^{-\beta \mathcal{H}_F(\tau)}e^{\beta \mathcal{H}(0)}\right],\label{g1}\\
&=&\text{Tr}_{\mathcal{G}}\Bigg[\rho_F(0)\mathcal{U}_{0,\tau}^{-1}\Theta^{-1}e^{\int_0^{\tau}\epsilon_A A_R(t)\lambda(\tau-t)dt}\Theta \mathcal{U}_{0,\tau}e^{-\beta \mathcal{H}_F(\tau)}e^{\beta \mathcal{H}(0)}\Bigg],\label{g2}\\
&=&Z(\beta,\lambda_i)^{-1}\text{Tr}_{\mathcal{G}}\Bigg[e^{\int_0^{\tau}\epsilon_A A_R(t)\lambda(\tau-t)dt}\Theta \mathcal{U}_{0,\tau}e^{-\beta \mathcal{H}_F(\tau)}\mathcal{U}_{0,\tau}^{-1}\Theta^{-1}\Bigg],\label{g3}\\
&=&Z(\beta,\lambda_i)^{-1}\text{Tr}_{\mathcal{G}}\Bigg[e^{\int_0^{\tau}\epsilon_A A_R(t)\lambda(\tau-t)dt}\Theta\mathcal{U}_{0,\tau}\mathcal{U}_{0,\tau}^{-1}e^{-\beta \mathcal{H}(\tau)}\mathcal{U}_{0,\tau}\mathcal{U}_{0,\tau}^{-1}\Theta^{-1}\Bigg],\nonumber\\  \label{g4}\\
&=&Z(\beta,\lambda_i)^{-1}\text{Tr}_{\mathcal{G}}\left[e^{\int_0^{\tau}\epsilon_A A_R(t)\lambda(\tau-t)dt}\Theta e^{-\beta \mathcal{H}(\tau)}\Theta^{-1}\right],\label{g5}\\
&=&\frac{Z(\beta,\lambda_f)}{Z(\beta,\lambda_i)}Z(\beta,\lambda_f)^{-1}\text{Tr}_{\mathcal{G}}\left[e^{\int_0^{\tau}\epsilon_A A_R(t)\lambda(\tau-t)dt}\Theta e^{-\beta \mathcal{H}(\tau)}\Theta^{-1}\right],\label{g6}\\
&=&\frac{Z(\beta,\lambda_f)}{Z(\beta,\lambda_i)}\text{Tr}_{\mathcal{G}}\left[e^{\int_0^{\tau}\epsilon_A A_R(t)\lambda(\tau-t)dt}\rho_R(0)\right],\label{g7}\\
&=&e^{-\beta\Delta F}\left\langle \exp\left(\int_0^{\tau}\epsilon_A A_R(t)\lambda(\tau-t)dt\right)\right\rangle_R. \label{g8}
\end{eqnarray}
From Equation \eqref{g1} to Equation \eqref{g2}, we have made use of Equation \eqref{d1} and \eqref{d2}. From Equation \eqref{g2} to Equation \eqref{g3}, we have taken advantage of cyclic property of trace operation, Equation \eqref{tr1}, and the definition of the initial state in the forward process, Equation \eqref{forward0}. From Equation \eqref{g3} to \eqref{g4}, we have made use of the Heisenberg representation of $\mathcal{H}_F(\tau)$. From \eqref{g6} to \eqref{g7}, we have made use of Equation \eqref{reversed1}. The last step comes from the definition of the expectation value in the reversed process. Thus the Theorem is proved. Now we show that some well known results can be recovered from the theorem proved above:\\
\textbf{1. Quantum Jarzynski equality for the $\mathcal{PT}$-symmetric quantum systems can be recovered from the quantum work relation for the $\mathcal{PT}$-symmetric quantum systems:}\\
If we take $\lambda(t)=0$, we then get an equality
\begin{eqnarray}
\left\langle e^{-\beta \mathcal{H}_F(\tau)}e^{\beta \mathcal{H}(0)}\right\rangle_F=e^{-\beta\Delta F}.
\end{eqnarray}
The left hand side can be understood as work measurement in the forward process because,
\begin{eqnarray}
&& \left\langle e^{-\beta \mathcal{H}_F(\tau)}e^{\beta \mathcal{H}(0)}\right\rangle_F\nonumber\\
&=&\text{Tr}_{\mathcal{G}}\left[\rho(0)e^{-\beta \mathcal{H}_F(\tau)}e^{\beta \mathcal{H}(0)}\right],\\
&=&\sum_{n}\langle n(\lambda_i)|G_0\left[\rho(0)\mathcal{U}_{0,\tau}^{-1}e^{-\beta \mathcal{H}(\tau)}\mathcal{U}_{0,\tau}e^{\beta \mathcal{H}(0)}\right]|n(\lambda_i)\rangle,\\
&=&\sum_{n,m}\langle n(\lambda_i)|G_0\rho(0)\mathcal{U}_{0,\tau}^{-1}e^{-\beta \mathcal{H}(\tau)}|m(\lambda_f)\rangle\nonumber\\
&&\times\langle m(\lambda_f)|G_{\tau}\mathcal{U}_{0,\tau}e^{\beta \mathcal{H}(0)}|n(\lambda_i)\rangle,\\
&=&\sum_{n,m}e^{-\beta(E_m(\lambda_f)-E_n(\lambda_i))}\langle n(\lambda_i)|G_0\rho(0)G_0^{-1}\mathcal{U}_{0,\tau}^{\dagger}G_{\tau}|m(\lambda_f)\rangle
\nonumber\\
&&\times\langle m(\lambda_f)|G_{\tau}\mathcal{U}_{0,\tau}|n(\lambda_i)\rangle,\\
&=&\sum_{n,m}Z(0)^{-1}e^{-\beta(E_m(\lambda_f)-E_n(\lambda_i))}\langle n(\lambda_i)|G_0e^{-\beta \mathcal{H}(0)}G_0^{-1}\mathcal{U}_{0,\tau}^{\dagger}G_{\tau}|m(\lambda_f)\rangle
\nonumber\\
&&\times\langle m(\lambda_f)|G_{\tau}\mathcal{U}_{0,\tau}|n(\lambda_i)\rangle,\\
&=&\sum_{n,m}Z(0)^{-1}e^{-\beta(E_m(\lambda_f)-E_n(\lambda_i))}\langle n(\lambda_i)|e^{-\beta \mathcal{H}(0)^{\dagger}}\mathcal{U}_{0,\tau}^{\dagger}G_{\tau}|m(\lambda_f)\rangle
\nonumber\\
&&\times\langle m(\lambda_f)|G_{\tau}\mathcal{U}_{0,\tau}|n(\lambda_i)\rangle,\\
&=&\sum_{n,m}p_n(\lambda_i)e^{-\beta(E_m(\lambda_f)-E_n(\lambda_i))}\langle n(\lambda_i)|\mathcal{U}_{0,\tau}^{\dagger}G_{\tau}|m(\lambda_f)\rangle
\nonumber\\
&&\times\langle m(\lambda_f)|G_f\mathcal{U}_{0,\tau}|n(\lambda_i)\rangle,\\
&=&\sum_{n,m}p_n(\lambda_i)e^{-\beta(E_m(\lambda_f)-E_n(\lambda_i))}\left|\langle m(\lambda_f)|G_{\tau}\mathcal{U}_{0,\tau}|n(\lambda_i)\rangle\right|^2,\\
&=&\int P(W)e^{-\beta W}dW,\\
&=&\langle e^{-\beta W}\rangle
\end{eqnarray}
Thus we recovered the quantum Jarzynski equality for the $\mathcal{PT}$-symmetric quantum systems \cite{PT2005} from the general work relation theorem we prove above.\\
\textbf{2. Response theory in $\mathcal{PT}$-symmetric quantum systems from the quantum work relation for the $\mathcal{PT}$-symmetric quantum systems:} Consider a perturbation of the form,
\begin{eqnarray}
\mathcal{H}(t)=\mathcal{H}_0-\eta(t)B.
\end{eqnarray}
Here $\eta(t)=0$ for $t\leq0,t\geq T$. The quantum observable $B$ is an arbitrary operator. To obtain the linear response of an arbitrary observable $A$ with respect to the perturbation $-\eta(t)B$, we perform functional derivative in the proved theorem with respect to $\lambda(\tau)$ around $\lambda=0$ and we get
\begin{eqnarray}\label{LR1}
\left\langle A_F(\tau) e^{-\beta \mathcal{H}_F(\tau)}e^{\beta \mathcal{H}(0)}\right\rangle_F&=&\epsilon_A\left\langle A_R(0)\right\rangle_R=\epsilon_A\left\langle A\right\rangle_R.
\end{eqnarray}
Here $\Delta F=0$ because $\eta(0)=\eta(\tau)=0$. Because the reversed process also begins at thermal equilibrium, the average on the right hand side of Equation \eqref{LR1} is an equilibrium average under time reversed Hamiltonian. However, according to time reversal symmetry, we have
\begin{eqnarray}\label{L1}
\epsilon_A\left\langle A\right\rangle_R=\langle A\rangle_F.
\end{eqnarray}
Note that this is true because the initial and final Hamiltonian in the linear response theory are the same. We are now to calculate the exponentials of the initial and final Hamiltonian. In the Heisenberg representation, the total time derivative of the Hamiltonian equals to its partial derivative and thus we have
\begin{eqnarray}
e^{-\beta \mathcal{H}_F(\tau)}&=&e^{-\beta(\mathcal{H}_0+\mathcal{H}_1)},
\end{eqnarray}
with
\begin{eqnarray}
\mathcal{H}_1&=&\int_0^{\tau}dt\left(\frac{\partial \mathcal{H}}{\partial t}\right)_F,\\
&=&-\int_0^{\tau}dt\dot{\eta}(t)B_F(t),\\
&=&\int_0^{\tau}dt\eta(t)\dot{B}_F(t).
\end{eqnarray}
In the last step above we have made use of the integral by parts. To go further, we make use of an exponential operator identity \cite{exponential},
\begin{eqnarray}
e^{\beta(\mathcal{P}+\mathcal{Q})}e^{-\beta \mathcal{P}}&=&1+\int_0^{\beta}du e^{u(\mathcal{P}+\mathcal{Q})}\mathcal{Q}e^{-u\mathcal{P}}.
\end{eqnarray}
To first order in $\mathcal{Q}$, we may neglect $\mathcal{Q}$ in the last exponential, $e^{u(\mathcal{P}+\mathcal{Q})}$. Taking $\mathcal{P}=-\mathcal{H}_0$ and $\mathcal{Q}=-\mathcal{H}_1$ and expanding to first order in $\eta$, we obtain
\begin{eqnarray}
&& e^{-\beta \mathcal{H}_F(\tau)}e^{\beta \mathcal{H}_0},\nonumber\\
&=&1-\int_0^{\tau}dt\eta(t)\int_0^{\beta}due^{-u\mathcal{H}_0}\dot{B}(t)e^{u\mathcal{H}_0}+o(\eta^2),\\
&=&1-\int_0^{\tau}dt\eta(t)\int_0^{\beta}du\dot{B}(t+iu)+o(\eta^2)\label{LR2}.
\end{eqnarray}
Here $B(t)=e^{it\mathcal{H}_0}Be^{-it\mathcal{H}_0}$. Inserting Equation \eqref{LR2} into the left hand side of Equation \eqref{LR1}, we get
\begin{eqnarray}\label{LR3}
\langle A_F(\tau)\rangle&=&\langle A\rangle+\int_0^{\tau}dt\eta(\tau-t)\phi_{AB}(t)+o(\eta^2).
\end{eqnarray}
Here the response function is defined as
\begin{eqnarray}\label{LR4}
\phi_{AB}(t)&=&\int_0^{\beta}du\langle\dot{B}(-iu)A(t)\rangle.
\end{eqnarray}
Equation \eqref{LR3} and Equation \eqref{LR4} are the expressions of linear response theory in the $\mathcal{PT}$-symmetric quantum system. \\
\textbf{3. Onsager reciprocity relations for the conductivity from the quantum work relation for the $\mathcal{PT}$-symmetric quantum systems:} Consider an electronic system and the current operator is
\begin{eqnarray}
\mathcal{J}_{\mu}=\sum_jq_j\dot{x}_{j,\mu},
\end{eqnarray}
where $q_j$ is the charge of the $j$-th electron and $x_{j,\mu}$ is the position of the $j$-th electron in the $\mu=x,y,z$ direction. If we consider a perturbation of the form $-\eta(t)\sum_jq_jx_{j,\nu}$, which is nonzero only within the time interval $0<t<\tau$ and zero otherwise. According to linear response theory, the response of the current in the $\mu$ direction is
\begin{eqnarray}
\langle \mathcal{J}_{\mu}(\tau)\rangle&=&\langle \mathcal{J}_{\mu}(0)\rangle+\int_0^{\tau}dt\eta(\tau-t)\phi_{\mu\nu}(t)+o(\eta^2),
\end{eqnarray}
where the response function is
\begin{eqnarray}
\phi_{\mu\nu}(t)&=&\int_{0}^{\beta}du\langle \mathcal{J}_{\nu}(-iu)\mathcal{J}_{\mu}(t)\rangle.
\end{eqnarray}
If the unperturbed Hamiltonian is time reversal invariant, we then have
\begin{eqnarray}
\phi_{\mu\nu}(t)&=&\int_{0}^{\beta}du\langle \mathcal{J}_{\nu}(-iu)\mathcal{J}_{\mu}(t)\rangle,\\
&=&Z(0)^{-1}\int_{0}^{\beta}du\text{Tr}_{\mathcal{G}}\left[e^{-\beta \mathcal{H}_0}\mathcal{J}_{\nu}(-iu)\mathcal{J}_{\mu}(t)\rangle\right],\label{ca1}\\
&=&Z(0)^{-1}\int_{0}^{\beta}du\text{Tr}_{\mathcal{G}}\left[e^{-\beta \tilde{\mathcal{H}}_0}\mathcal{J}_{\nu}(-iu)\mathcal{J}_{\mu}(-t)\rangle\right],\label{ca2}\\
&=&Z(0)^{-1}\int_{0}^{\beta}du\text{Tr}_{\mathcal{G}}\left[e^{-\beta \tilde{\mathcal{H}}_0}\mathcal{J}_{\nu}(t)\mathcal{J}_{\mu}(iu)\rangle\right]\label{ca3},\\
&=&Z(0)^{-1}\int_{0}^{\beta}du\text{Tr}_{\mathcal{G}}\left[e^{-\beta \tilde{\mathcal{H}}_0}\mathcal{J}_{\mu}(-iu)\mathcal{J}_{\nu}(t)\rangle\right]\label{ca4},\\
&=&\phi_{\nu\mu}(t).
\end{eqnarray}
From Equation \eqref{ca1} to Equation \eqref{ca2}, we have made use of the time reversal symmetry, where $\tilde{\mathcal{H}}_0$ is the time reversed Hamiltonian of $\mathcal{H}_0$. From Equation \eqref{ca2} to Equation \eqref{ca3}, we just rearrange the exponential factors. From Equation \eqref{ca3} to Equation \eqref{ca4}, we have made use of the Kubo-Martin-Schwinger formula \cite{Kubo1957,KMS1,KMS2}. The conductivity is the Fourier transform of the response function, namely $\sigma_{\mu,\nu}(\omega)=\int_0^{\infty}dte^{i\omega t}\phi_{\mu,\nu}(t)$, thus we have
\begin{eqnarray}
\sigma_{\mu,\nu}(\omega)&=&\sigma_{\nu,\mu}(\omega).
\end{eqnarray}
We thus derived the Onsager reciprocal relations \cite{Onsager1931a,Onsager1931b,Casimir1945} for the conductivity in $\mathcal{PT}$-symmetry quantum systems.

Finally we make a brief remark on the case when the Hamiltonian $\mathcal{H}(t)$ is in the regime of broken $\mathcal{PT}$ symmetry. In this regime, the eigenvalues have both real and imaginary parts so that the quantum dynamics is non-unitary. Because our derivations for the quantum work relations rely on the unitarity of dynamics (see the Lemma), we may conclude that the quantum work relation does not hold in the regime of broken $\mathcal{PT}$ symmetry.

\section{Conclusions}  We have demonstrated that a universal quantum work relation for a system driven arbitrarily far from equilibrium generalized to the $\mathcal{PT}$-symmetric quantum systems which is in the phase of unbroken $\mathcal{PT}$ symmetry. This relation is a consequence of microscopic reversibility for non-equilibrium thermodynamics. We have recovered the quantum Jarzynski equality, linear response theory and the Onsager reciprocal relations for the $\mathcal{PT}$-symmetric quantum system as special cases of the universal quantum work relation in $\mathcal{PT}$-symmetric quantum systems. In the regime of broken $\mathcal{PT}$ symmetry, the universal quantum work relation does not hold as the norm is not preserved during the dynamics.

\begin{acknowledgements}
This work was supported by National Natural Science Foundation of China (Grants No. 11604220) and the Start Up Fund of Shenzhen University (Grants No. 2016018).
\end{acknowledgements}

\end{document}